\documentstyle[12pt]{article}
\begin{document}
\begin{titlepage}
\begin{flushright}
BUTP-95/21
\end{flushright}
\vspace{0.2in}
\begin{center}
{\Large \bf On the mass splittinq between axial and vector
heavy-light mesons \\ }
\vspace{0.4in} {\bf V.L. Eletsky}
\\ Institute of Theoretical and
Experimental Physics \\ B. Cheremushkinskaya 25, 117259 Moscow, Russia \\
\vspace{0.2in}
and \\
\vspace{0.2in}
Institute for Theoretical Physics, Bern University \\
Sidlerstrasse 5, CH-3012 Bern, Switzerland \\
\vspace{0.4in}
{\bf   Abstract  \\ }
\vspace{0.2in}
\end{center}

Mass splitting between axial and vector $\bar{Q}q$ mesons is considered
within the standard QCD sum rules. In agreement with the first
experimental data on the $B_1$ meson ($J^P =1^{+}$) we find that the
splitting for B is about the same as for D and show that $1/m_Q$ corrections
to the meson masses are small.

\end{titlepage}

Several groups\cite{exp} recently
announced that they had observed candidates for the axial open beauty state
($J^P_{j} =1^{+}_{3/2}$) about $500$ MeV above the corresponding vector
meson $B^{\ast}(5325)$. The splitting appears to be of the same magnitude as
in the case of D mesons, as predicted e.g. in the instanton liquid
model\cite{s2}.
This fact however looks somewhat strange from the
viewpoint of expansion of the meson mass with respect to $1/m_Q$, where
$m_Q$ is the heavy quark mass,

\begin{equation}
m=m_Q + E_0 +\frac{E_1}{m_Q}+...,
\label{mass}
\end{equation}
and indicates that the $1/m_Q$ corrections are either very close in both
the vector and axial cases, or small. On one hand, there appears to be no
reason for these corrections to be close (e.g., the resonance energy, $E_0$,
is different in both cases\cite{s1}). On the other hand, it is known that
for the couplings, such as $f_D$ and $f_B$, these corrections are very
important, and $1/m_Q$-expansion breaks down for $f_D$\cite{ae,es}.
In the present paper we calculate the mass of the axial $B_1$ meson
and demonstrate, using a non-relativistic version\cite{s1,es} of the standard
QCD sum rules\cite{svz}, that $1/m_Q$ corrections to the $\bar{Q}q$ vector
and axial meson masses are indeed rather small.

The sum rules for vector and axial $\bar{Q}q$ mesons are obtained by
considering correlators of vector ($j_{\mu}=\bar{Q}\gamma_{\mu}q$)
and axial ($j_{\mu}=\bar{Q}\gamma_{\mu}\gamma_5 q$)
currents
\begin{equation} C_{\mu\nu}(q)=i\int d^{4}x e^{iqx} \langle
0|T\{j_{\mu}(x),j_{\nu}^{+}(0)|0\rangle \label{c}
\end{equation}
at Euclidean momentum $-q^2 > 1\,$ GeV$^2$. We choose the tensor structure
proportional to $g_{\mu\nu}$, since for $q_{\mu}q_{\nu}$ the lowest state in
the axial case is the pseudoscalar meson, and take into account operators
with dimension $d\le 5$ omitting the gluon condensate whose contribution is
negligible.
After the standard Borel transformation the sum rule takes the
form\cite{ek,be}

\begin{eqnarray}
\lefteqn{\frac{m^{4}_{\pm}}{g^{2}_{\pm}}\exp (-m_{\pm}^2/M^2)}
\nonumber\\
& &=\frac{M^4}{8\pi^2}\int_{m_Q^2}^{s_0}ds s e^{-s/M^2}
\left( 1-\frac{m_Q^2}{s}\right)^2\left( 2+\frac{m_Q^2}{s}\right)
\left( 1+\frac{4\alpha_s (M^2)}{3\pi}F(\frac{m_Q^2}{s})\right)
\nonumber\\
& & \pm m_Q\langle\bar{q}q\rangle L^{4/9}\exp (-m_Q^2/M^2)
\mp\frac{1}{4M^4}m_{Q}^{3}m_{0}^2\langle\bar{q}q\rangle\exp (-m_Q^2/M^2)
\label{rel}\; .
\end{eqnarray}
The couplings $g_{\pm}$ are defined according to
$\langle 0|j_{\mu}|1^{\pm}\rangle =(m^{2}_{\pm}/g_{\pm})e_{\mu}$,
where $e_{\mu}$ is the meson polarization vector. The function $F(x)$
decribes the $\alpha_s$-corrections\cite{rad},

\begin{eqnarray}
F(x)&=&\frac{13}{4}+2Li_{2}(x)+\ln x \ln (1-x) +\frac{3}{2}\frac{x}{2+x}
\ln\left(\frac{x}{1-x}\right)
\nonumber\\
&-&\ln (1-x)-\frac{4-x-x^2}{(1-x)^2 (2+x)}x \ln x -
\frac{5-x-2x^2}{(1-x)(2+x)}\; .
\label{rad}
\end{eqnarray}
The other quantities in Eq.(\ref{rel}) are:
$m_0^2 =\langle\bar{q}\sigma_{\alpha\beta}G_{\alpha\beta}q\rangle
/\langle\bar{q}q\rangle =0.8$\, GeV$^2$, $\langle\bar{q}q\rangle =
-(0.24)^3$\, GeV$^3$, $L=\ln (M/\Lambda)/\ln (\bar{\mu}/\Lambda)$, $\Lambda =
0.15$ GeV, $\bar{\mu} =0.5$\, GeV.  The meson mass is obtained from
Eq.(\ref{rel}) by taking the logarithmic derivative with respect to $M^2$.
The only difference between the vector and axial sum rule (apart from the
value of the continuum threshold, $s_0$) is
the sign of the terms with $\langle\bar{q}q\rangle$. Neglecting the continuum
and the anomalous dimension factor, it is easy to obtain the following
estimate

\begin{equation}
\frac{m^2_{+}-m^2_{-}}{m^2_{+}+m^2_{-}}\approx
-\frac{4\pi^2 m_Q\langle\bar{q}q\rangle}{M^4}
\left( 1-\frac{5m_0^2 m_Q^2}{12M^4}\right)\; .
\label{dm2}
\end{equation}
In the limit $m_Q\to\infty$
the Borel parameter scales as $M^2\sim 2m_Q\mu$, where $\mu=1/\tau$ is the
non-relativistic Borel parameter and $\tau$ is the typical euclidean
time over which the correlators change
significantly\cite{es}. Thus, one expects for the splitting between
axial and vector $\bar{Q}q$ states at $m_Q\to\infty$

\begin{equation}
m_{+}-m_{-}\approx -\frac{\pi^2\langle\bar{q}{q}\rangle}{\mu^2}
\left( 1-\frac{5m_0^2}{48\mu^2}\right)\; .
\label{dm}
\end{equation}
The r.h.s. of the above equation has a flat maximum at $\mu\approx 0.4$ GeV
which corresponds to $m_{+}-m_{-}\approx 0.4$ GeV.
The question is whether $b$ and $c$ quarks are heavy enough for B and D
mesons to satisfy Eq.(\ref{dm}), i.e. whether $1/m_Q$-corrections to
masses are really small. Our experience with the couplings\cite{es}
indicates that these corrections may be important.

In Figs. 1 and 2 we show the results of numerical analysis of the sum rules
of Eq.(\ref{rel}) in the corresponding working windows in $M^2$ for
$m_c=1.35$ GeV, $m_b =4.7$ GeV and the optimal values of continuum
thresholds $s_0 = 6,\; 8,\; 35$ and $40$ GeV$^2$ for $D^{\ast}$,
$D_1$, $B^{\ast}$ and $B_1$, respectively\cite{ek,be,b}.
We see that the mass splitting is indeed
the same for D and B:

\begin{equation}
m_{+}-m_{-}=(500\pm 50)\, MeV\; .
\label{numdm}
\end{equation}
For the masses, the sum rules give the values $m_{B_1}=(5.9\pm 0.15)$ GeV,
$m_{B^{\ast}}=(5.4\pm 0.15)$ GeV, $m_{D_1}=(2.5\pm 0.15)$ GeV and
$m_{D^{\ast}}=(2.0\pm 0.15)$ GeV. The errors correspond to allowed
variations of continuum threshold $s_0$. A better accuracy for the splitting
is due to partial cancelation of continuum contributions in this case. For
the couplings we get $g_{B_1}=22\pm 3$, $g_{B^{\ast}}=27\pm 3$,
$g_{D_1}=10.5\pm 1.5$ and $g_{D^{\ast}}=9\pm 1.5$ (the last two couplings
were obtained before\cite{ek,be}). It is worth mentioning that
$\alpha_s$-corrections are rather important: without them the splittings
would be about $30\%$ bigger.

Now, we will explicitly demonstrate that the $1/m_Q$-correction to the
meson mass in Eq.(\ref{mass}) is small by expanding the sum rule of
Eq.(\ref{rel}) around the limit $m_Q\to\infty$. Following the
same procedure as in\cite{es}, we introduce the non-relativistic continuum
threshold $E_c$ and Borel parameter $\mu$, $s_0=(m_Q+E_c)^2$, $M^2=2m_Q\mu$,
and obtain

\begin{equation}
E_0 =3\mu\frac{F_2(E_c/\mu)a_2^{rad}\pm\pi^2
m_0^2\langle\bar{q}q\rangle/144\mu^5 }
{F_1(E_c/\mu)a_1^{rad}\pm\pi^2\langle\bar{q}q\rangle L^{4/9}/6\mu^3\mp
\pi^2 m_0^2\langle\bar{q}q\rangle/96\mu^5}\; ,
\label{e0}
\end{equation}

\begin{eqnarray}
E_1 =&-&\frac{1}{2}E_0^2 -\frac{48}{C_0\pi^2}\exp(E_0/\mu)\mu^4
(4\mu-E_0)F_2 (E_c/\mu)a_2^{rad}  \nonumber\\
&-& \frac{3}{2C_0\pi^2}\exp[(E_0-E_c)/\mu]E_c^4
[(E_0-E_c)a_1^{rad}-(11\mu /2)a_2^{rad}]\; .
\label{e1}
\end{eqnarray}
where $C_0$ determines the asymptotic behavior of the residue
in the limit $m_Q\to\infty$,

\begin{equation}
\frac{1}{g^2}=\frac{C_0}{m_Q^3}\left( 1+\frac{C_1}{m_Q}+ ...\right)\; .
\label{g}
\end{equation}
The coefficient $C_0$ itself is determined from the following sum rule

\begin{equation}
C_0 = \exp (E_0/\mu)\left(\frac{6\mu^3}{\pi^2}F_1(E_c/\mu)a_1^{rad}
\pm\langle\bar{q}q\rangle L^{4/9} \mp m_0^2 \langle\bar{q}q\rangle /16\mu^3
\right)\; .
\label{c0}
\end{equation}
Upper and lower signs in Eqs.(\ref{e0}) and (\ref{c0})
correspond to axial and vector cases, respectively.
Once $E_1$ is known, $1/m_Q$-correction to the coupling can be obtained from
the sum rule

\begin{eqnarray}
C_1 =&-&4E_0 +\frac{E_0^2+2E_1}{2\mu}-\frac{48}{C_0\pi^2}\exp(E_0/\mu)
\mu^4 F_2 (E_c/\mu)a_2^{rad} \nonumber\\
&+& \frac{3}{2C_0 \pi^2}\exp[(E_0-E_c)/\mu]E_c^4 a_1^{rad}\; .
\label{c1}
\end{eqnarray}
The functions $F_1$ and $F_2$ in Eqs.(\ref{e0}), (\ref{e1}), (\ref{c0})
and (\ref{c1}) are the standard functions describing quark loop and continuum
contributions,

\begin{eqnarray}
F_1(x) &=& 1-(1+x+x^2/2)e^{-x}\, , \nonumber\\
F_2(x) &=& 1-(1+x+x^2/2+x^3/6)e^{-x}\; .
\label{F12}
\end{eqnarray}
The factors $a_1^{rad}$ and $a_2^{rad}$ contain radiative
corrections
\footnote{As in ref.\cite{es}, the arguments of the logs in $a_{1,2}^{rad}$
correspond to the maxima of the integrands in the dispersion integrals.
In the numerical analysis we put $m_Q = m_b$. Thus, we do not
actually go to the limit $m_Q\to\infty$ in the radiative corrections. For
a rigorous treatment of radiative corrections in this limit within the heavy
quark effective theory, see e.g.\cite{br}.},

\begin{eqnarray}
a_1^{rad}&=&1+\alpha_s \left( 2.35+\frac{2}{\pi}\log (m_Q/4\mu)\right)
\; , \nonumber\\
a_2^{rad}&=&1+\alpha_s \left( 2.35+\frac{2}{\pi}\log (m_Q/6\mu)\right)
\label{a12}\; .
\end{eqnarray}
Note, that in the leading order in $1/m_Q$ the sum rules for $J=1$ and
$J=0$ meson masses\cite{s1} are the same. The hyperfine splitting is
contained in the $1/m_Q$-corrections.

The asymptotic coefficients $E_0$ and $C_0$ were calculated in ref.\cite{s1},
and the splitting $E_0^{+} - E_0^{-}=800\pm 200$ MeV was obtained.
Our results for $E_0^{+}$ and $E_0^{-}$ are presented in Fig.3 for the
non-relativistic continuum thresholds $E_c =1.8$ GeV and $E_c =1$ GeV in
the axial and vector cases respectively. We obtained a smaller value,
$E_0^{+} - E_0^{-}=600\pm 100$ MeV.  We were not able to trace the source of
this difference, since the axial case was discussed very briefly in
ref.\cite{s1}.  For $C_0^{\pm}$ we then get $C_0^{+}\approx 0.71$ and
$C_0^{-}\approx 0.16$ (Fig.4) which agrees with ref.\cite{s1}
\footnote{This value of $C_0^{-}$ is in agreement with the estimate
obtained in an earlier paper\cite{dp}.}.
Using these values in Eq.(\ref{e1})
we obtain the results for $1/m_Q$  mass corrections presented in Fig.5. We
see that the stability of the sum rule for $E_1$ and the accuracy is rather
poor in the axial case.  However, it is clear that the $1/m_Q$-corrections
turn out small both on the scale of $m_b$ and $m_c$
\footnote{$1/m_Q$ corrections to vector and pseudoscalar meson masses were
calculated within the sum rules approach in Refs.\cite{n} and \cite{bb}.
Our value for $E_1$ in the vector case is of the same sign as in\cite{n},
but four times smaller in the absolute value. On the other hand, it is two
times smaler in the absolute value and of the opposite sign than
in\cite{bb}. The disagreement is disturbing and should be resolved. But we
do not discuss it here, since the corrections are small in any case.}.
Finally, using the values $E_1^{+}=0$ and $E_1^{-}=-0.12$ GeV$^2$ we get from
Eq.(\ref{c1}) $C_1^{+}\approx -2.6$ GeV and $C_1^{-}\approx - 5.5$ GeV
(Fig.6).

Thus, we have shown that the splittings between axial and vector
mesons with open charm and beauty calculated from QCD sum rules are
rather close and agree with experiment. By a non-relativistic expansion of
the sum rules we checked that $1/m_Q$-corrections to meson masses are very
small.  This is in contrast with similar corrections to the couplings which
are very important (in the vector case $1/m_Q$ expansion does not work for
$g_{B^{\ast}}$!).

I am grateful to A.B. Kaidalov for asking a question which triggered
this calculation. I am indebted to H. Leutwyler for the warm hospitality at
the University Bern where this work has been done.  This work was supported
in part by Schweizerischer Nationalfonds and by the INTAS grant 93-0283.

\pagebreak
{\large\bf Figure Captions:}
\begin{itemize}
\item
Fig. 1: Mass splitting between $D_1$ and $D^{\ast}$ from (\ref{rel}) in GeV.
\item
Fig. 2: Same as Fig. 1 for $B_1$ and $B^{\ast}$.
\item
Fig. 3: $E_0$ in GeV.
\item
Fig. 4: $C_0$ in GeV$^3$.
\item
Fig. 5: $E_1$ in GeV$^2$.
\item
Fig. 6: $C_1$ in GeV.
\end{itemize}

\newpage
\begin{thebibliography} {99}
\bibitem{exp} OPAL Collaboration, R. Akers et al, Z.Phys. C {\bf 66},
              19 (1995); \\
              DELPHI Collaboration, P.Abreu et al, Phys. Lett.
              B {\bf 345}, 598 (1995); \\
              ALEPH Collaboration, presented by S. Schael at Moriond
              Workshop, March, 1995.
\bibitem{s2}  E.V. Shuryak, Nucl. Phys. {\bf B328}, 85 (1992).
\bibitem{s1}  E.V. Shuryak, Nucl. Phys. {\bf B198}, 83 (1982).
\bibitem{ae}  T.M. Aliev and V.L. Eletsky, Sov. J. Nucl. Phys.
              {\bf 38}, 936 (1983).
\bibitem{es}  V.L. Eletsky and E.V. Shuryak, Phys. Lett. B {\bf 276},
              191 (1992).
\bibitem{svz} M.A. Shifman, A.I. Vainshtein and V.I. Zakharov,
              Nucl. Phys. B {\bf 147}, 385, 448, 519 (1979).
\bibitem{ek}  V.L. Eletsky and Ya.I. Kogan, Z.Phys. C {\bf 28},
              155 (1985).
\bibitem{be}  B.Yu. Blok and V.L. Eletsky, Sov. J. Nucl. Phys.
              {\bf 42}, 787 (1985).
\bibitem{rad} D.J. Broadhurst, Phys. Lett. B {\bf 101}, 423 (1981).
\bibitem{b}   V.M. Belyaev, V.M. Braun, A. Khodjamirian and R. R\" uckl,
              Phys. Rev. {\bf D 51}, 6177 (1995).
\bibitem{br}  E. Bagan, P.Ball, V.M. Braun and H.G. Dosch, Phys. Lett.
              B {\bf 278}, 457 (1992).
\bibitem{dp}  C.A. Dominguez and N. Paver, Phys. Lett. B {\bf 246}, 493
              (1990).
\bibitem{n}   M. Neubert, Phys.Rev. {\bf D 45}, 2451 (1992).
\bibitem{bb}  P. Ball and V.M. Braun, Phys. Rev. {\bf D 49}, 2472 (1994).
\end{thebibliography}
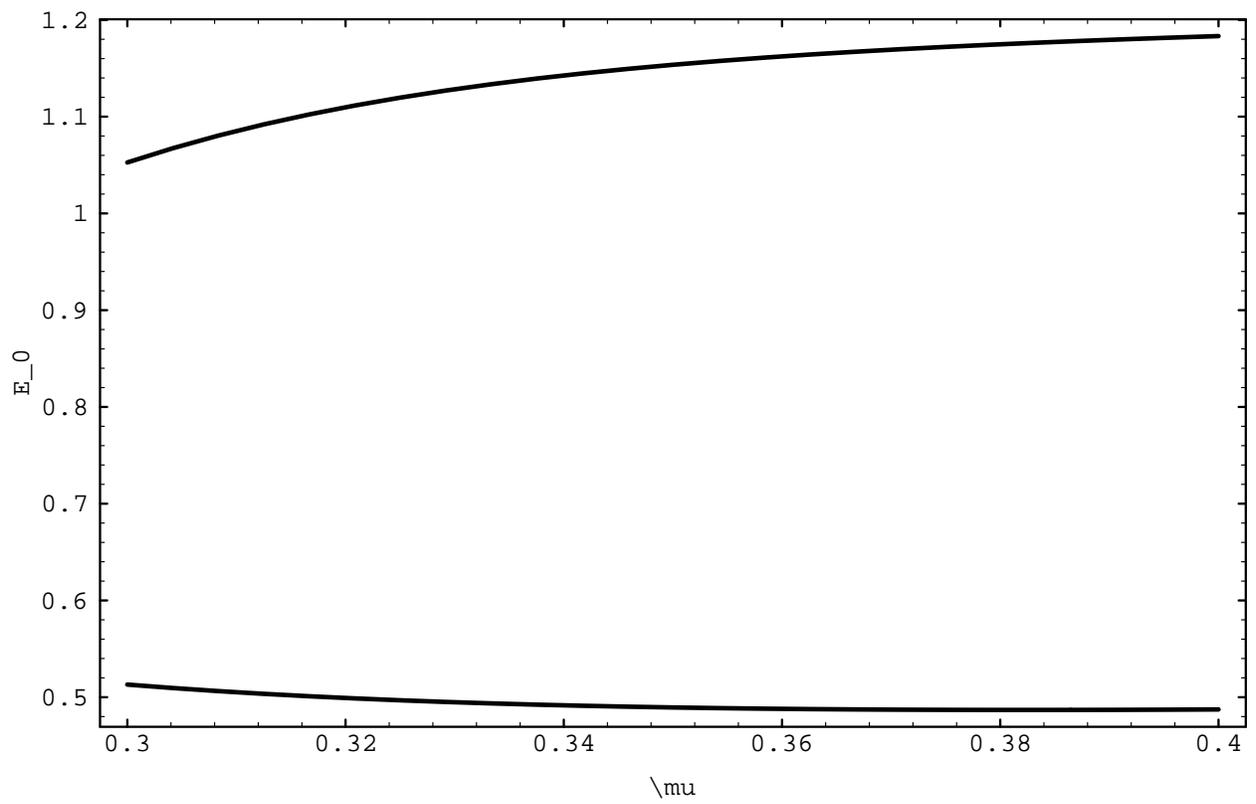
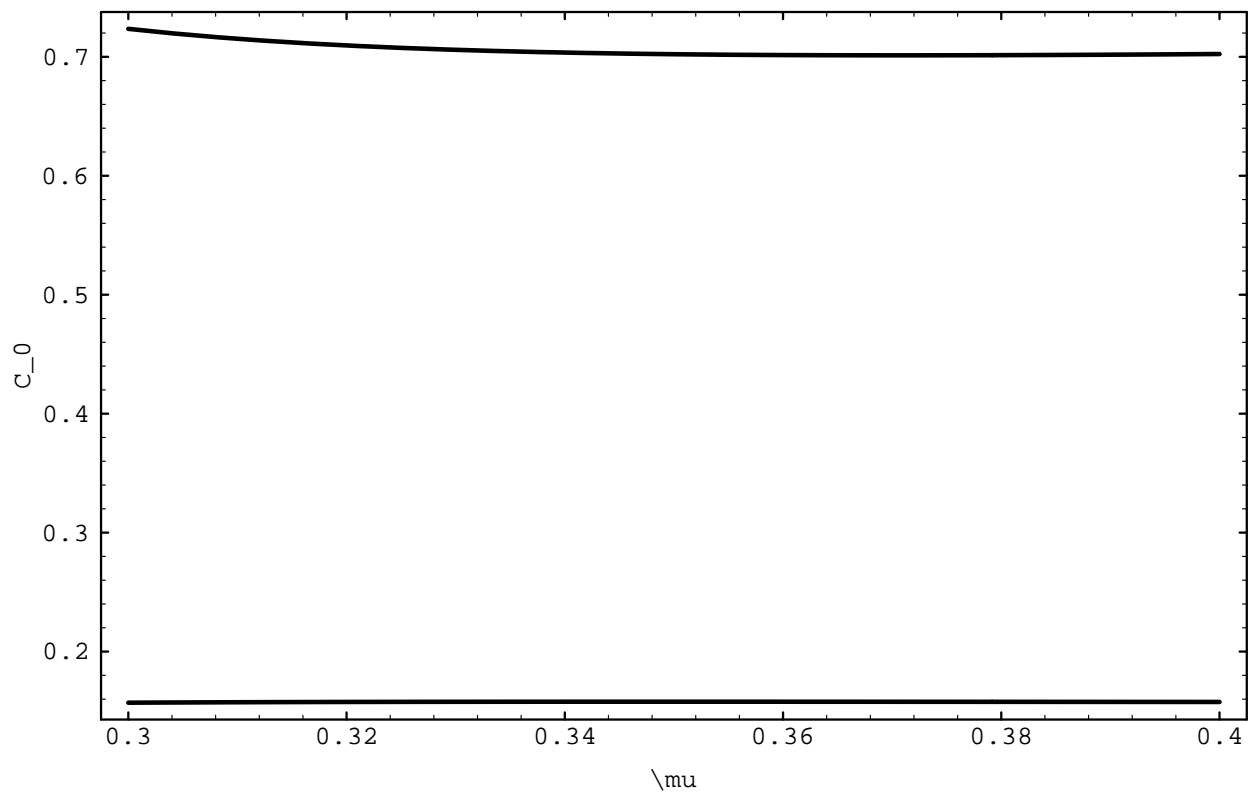
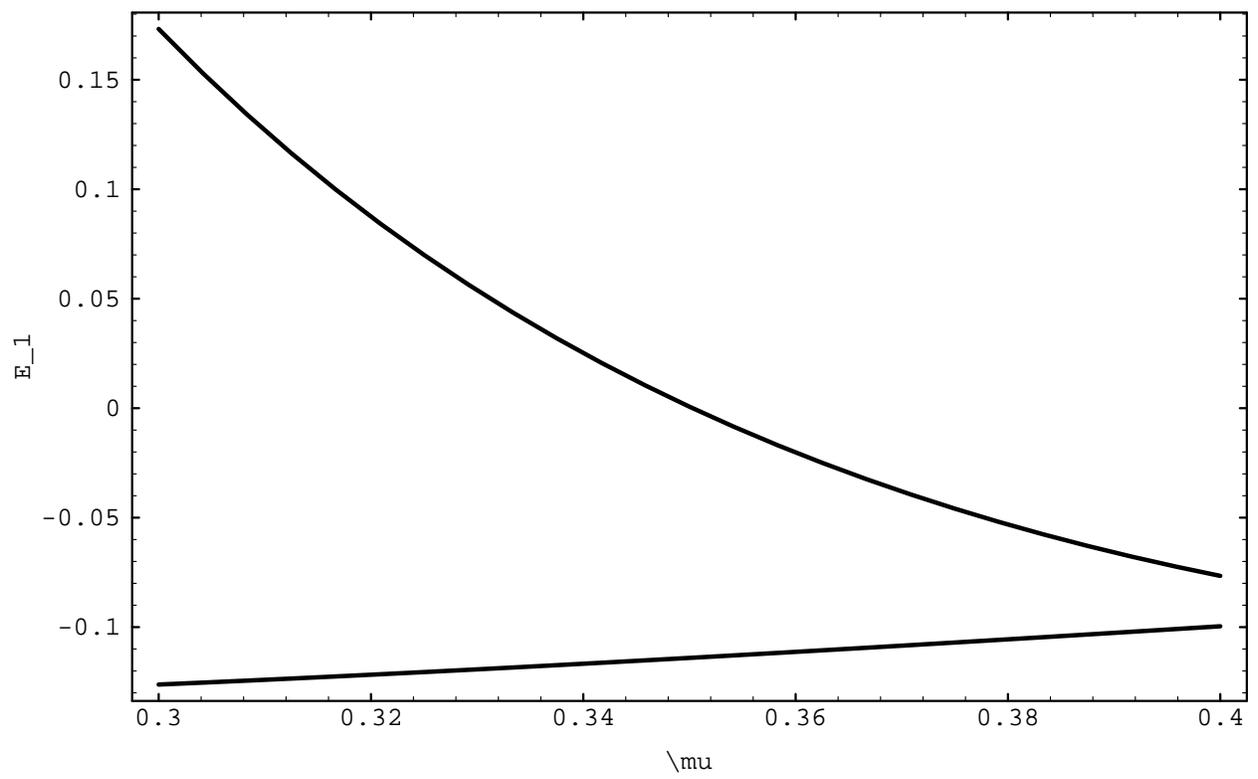
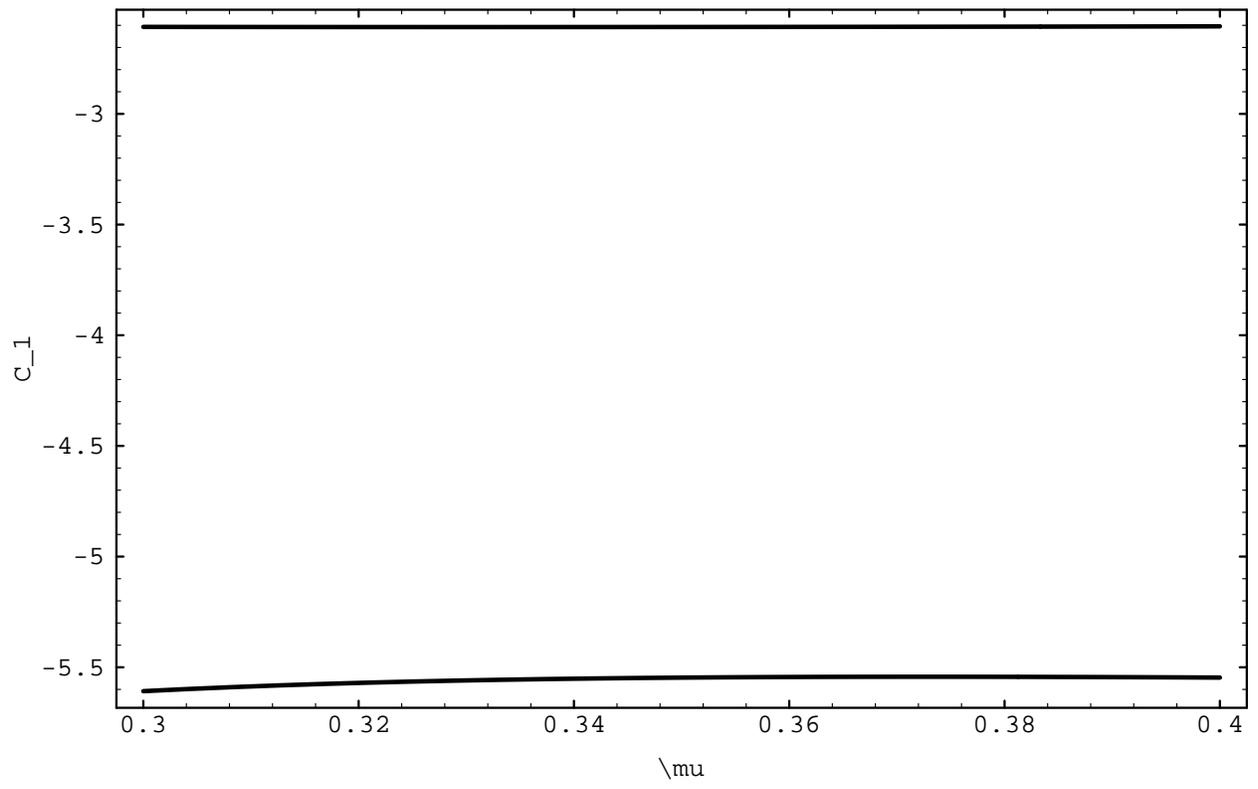
\end{document}